\documentclass[journal]{IEEEtran}
\IEEEoverridecommandlockouts

\usepackage{cite}
\usepackage{amsmath,amssymb,amsfonts}
\usepackage{amsthm}
\usepackage{algorithmic}
\usepackage{graphicx}
\usepackage{textcomp}
\usepackage{xcolor}
\usepackage{bm}
\usepackage{multirow}
\usepackage{comment}
\usepackage[caption=false,font=normalsize,labelfont=sf,textfont=sf]{subfig}

\renewcommand{\H}{\mathsf{H}}
\newcommand{\T}{\mathsf{T}}

\newcommand{\fl}{\mathrm{fl}}
\newcommand{\fd}{\mathrm{fd}}
\newcommand{\cc}{\mathrm{cc}}
\newtheorem{theorem}{Theorem}
\newtheorem{lemma}[theorem]{Lemma}
\newtheorem{proposition}[theorem]{Proposition}
\newtheorem{definition}[theorem]{Definition}

\begin{document}

\title{Wideband Large-Array Processing and Sparse Design for Angle Imaging
}

\author{
        Ziyu~Zhou and Wei~Dai%
        \thanks{This work has been submitted to the IEEE for possible publication. Copyright may be transferred without notice, after which this version may no longer be accessible.}
        \thanks{A preliminary version of this work was published in the Proceedings of the 2026 IEEE International Conference on Acoustics, Speech and Signal Processing (ICASSP)~\cite{11462930}. The present paper extends the coverage criterion, its theoretical analysis, and the associated sparse-array design to frequency-dependent channel responses, and reformulates the stability analysis from a probabilistic characterization to deterministic condition-number bounds.}%
        \thanks{Ziyu Zhou and Wei Dai are with the Department of Electrical and Electronic Engineering, Imperial College London, London, UK (e-mail: ziyu.zhou20@imperial.ac.uk; wei.dai1@imperial.ac.uk).}%
}

\maketitle

\begin{abstract}

This paper shows that wideband large-array processing can recover a large number of angle pixels with far fewer antenna elements. 
The key advantage of wideband signaling is that different frequencies induce different virtual arrays, whose union forms a virtual array with a substantially increased number of effective virtual elements. 
Thus, a sparse physical array can support far more spatial samples than physical antennas.
Motivated by this capability, we study the recovery of angular responses across the full field of view $[-90^\circ, 90^\circ)$, discretized according to the improved angular resolution, and refer to this sensing regime as angle imaging.

However, the resulting virtual array is inherently irregular, clustered, and does not automatically guarantee stable recovery. 
To address this challenge, we introduce a coverage criterion that estimates the number of stably recoverable angle pixels, without computationally intensive singular-value-based conditioning tests over candidate image dimensions. 
For systems satisfying this criterion, we theoretically establish deterministic condition-number bounds that characterize stable angle imaging.
Building on this criterion, we derive non-uniform sparse array designs that minimize the number of physical antennas while maintaining recovery over the full field of view. 
Simulation results show that the proposed criterion provides practical guidance for stable system design, and that the resulting sparse arrays can recover substantially more angle pixels than the number of physical antennas, with representative designs supporting over ten times as many angle pixels as physical antennas.
\end{abstract}


\section{Introduction}

High-resolution angular sensing is becoming increasingly important in millimeter-wave (mmWave), sub-terahertz, and integrated sensing and communication systems, where large bandwidths and large apertures are expected to support environment perception, target localization, and spatial channel acquisition \cite{heath2016overview,liu2020joint,liu2022isac,zhang2022jcas}. In these applications, extended objects and multiple nearby reflectors may occupy a continuum of directions rather than a small set of isolated angles. Therefore, it is desirable to reconstruct an angular response over a prescribed field of view (FoV), so that extended objects and distributed reflectors can be represented on a dense angular grid. We refer to this sensing regime as angle imaging. For such a task, the key performance requirements are angular resolution, numerical stability, and hardware cost.

Conventional angle imaging approaches face a high cost when fine angular resolution and stable recovery are required.
A standard way to form an angle image is to acquire antenna-domain measurements with fully digital arrays and apply matched filtering or Fourier beamforming over a prescribed angular grid \cite{fan20244d,barton1980digital}. 
These conventional beamforming formulations are typically developed under a narrowband signal model, where avoiding spatial ambiguity over the FoV requires half-wavelength antenna spacing. 
Thus, increasing angular resolution by enlarging the aperture also increases the number of antennas and RF chains \cite{wu2024real}. 
Analog and hybrid arrays reduce the radio-frequency (RF)-chain cost by observing weighted sums of antenna outputs, but multiple beam scans are required to cover the angular domain \cite{mandelli2022sampling}. 
Capon/minimum-variance distortionless response (MVDR) beamforming \cite{capon2005high, oh2015low, dmochowski2008linearly} can further improve the angular spectrum using covariance information, but additional snapshots and computational requirements are introduced. 
Therefore, conventional angle imaging faces an inherent tension among angular resolution, numerical stability, and hardware or acquisition cost.

Closely related methods, such as sparse-array and super-resolution methods, can reduce hardware cost or improve angular localization, while they mainly target sparse angular recovery rather than angle imaging.
Sparse arrays, such as minimum-redundancy arrays \cite{moffet1968minimum}, nested arrays \cite{pal2010nested, liu2016super}, coprime arrays \cite{vaidyanathan2011sparse,tan2014coprime,qin2015generalized}, and low-redundancy symmetric arrays 
\cite{rajamaki2021sparse}, increase covariance-domain degrees of freedom using fewer physical antennas. Compared with fully populated digital arrays of comparable aperture, these designs reduce the RF chain cost \cite{chen2008minimum, rajamaki2020hybrid, sun20214d, mukherjee2023compact}. 
Subspace-based parametric methods, such as MUSIC \cite{schmidt1986multiple} and ESPRIT \cite{roy1989esprit}, together with sparse-reconstruction methods, including on-grid \cite{malioutov2005sparse,stoica2011spice} and off-grid compressed sensing \cite{yang2013offgrid,tang2013compressed,bhaskar2013atomic,yang2015gridless,wu2023gridless,wu2024coffee}, can localize angular components with resolution finer than conventional beamforming under sparsity, separation, or sufficient signal-to-noise ratio (SNR) assumptions.
Nevertheless, a framework that jointly provides fine angular resolution, numerical-stability guidance, and reduced hardware cost for angle imaging without relying on angular sparsity assumptions remains underdeveloped.

Wideband array processing offers a new opportunity for hardware-efficient angle imaging from a single snapshot without relying on angular sparsity assumptions.
This opportunity arises from frequency diversity, as different frequency components induce distinct spatial steering phases across the same physical aperture.
This spatial-wideband behavior, often discussed through beam squint and related dual-wideband effects, has usually been treated as a modeling challenge for channel estimation and beamforming \cite{wang2018spatial,wang2019beam, 9646498}. 
However, the same phenomenon also creates an opportunity for angle imaging \cite{9771341,10058989,10271123}. 
Each frequency-antenna pair can be interpreted as a virtual spatial sample. 
Consequently, the union of all frequency-induced samples forms a composite virtual array whose number of effective spatial samples can be much larger than the number of physical antennas. 
When this composite virtual array is properly distributed, it can reduce spatial ambiguities, relax the need for half-wavelength physical antenna spacing, and enable sparse physical arrays to support dense angle imaging.

This opportunity also introduces a new stability problem. 
The virtual arrays generated by frequency diversity are inherently irregular and often clustered over the aperture \cite{54811}. 
Even when the total number of virtual elements is large, large gaps may remain between adjacent virtual elements. 
Such gaps can cause strong correlations among columns of the angle-imaging matrix, which leads to unstable recovery. This difficulty becomes more pronounced when the scattering response is frequency-dependent \cite{wang2019beam}, as the measurements must then distinguish not only different angular responses, but also different frequency-variation patterns. 

To address this challenge, this paper develops a geometry-driven framework for stable wideband angle imaging. The central idea is to link the numerical stability, traditionally assessed through explicit condition-number evaluation, to the geometric pattern of the virtual array. Based on this link, we introduce a coverage criterion (CC), which determines the maximum number of contiguous angle pixels stably supported by a given physical array and signal band. Specifically, the CC evaluates whether the frequency-induced virtual elements provide full coverage for the desired image dimension. In this way, the abstract conditioning requirement is converted into a straightforward geometric test, which further guides sparse-array design for stable full-FoV angle imaging.

\subsection{Our Contributions}

This paper develops a coverage-based framework for stable angle imaging with sparse physical arrays. The proposed framework jointly achieves fine angular resolution, numerical stability, and reduced physical-array cost. The main contributions are summarized as follows.

\begin{itemize}
    \item \textit{Coverage criterion and deterministic stability guarantee:}
    We develop the CC that characterizes whether a physical array and signal band can stably support a prescribed number of contiguous angle pixels. 
    By linking the largest uncovered gap of the virtual aperture to the conditioning of the imaging matrix, we prove a deterministic condition-number bound for CC-constructed systems. 
    This converts the stability requirement of wideband angle imaging into a straightforward geometric test.

    \item \textit{CC-guided closed-form sparse-array design:}
    We derive a closed-form non-uniform sparse-array design for stable full-FoV angle imaging with much fewer physical antennas. 
    The resulting antenna positions are obtained directly from the virtual-aperture coverage requirement, avoiding iterative or combinatorial array search. This provides a computationally efficient design procedure that reduces physical-array cost while maintaining stable recovery.
    Numerical results verify that the designed arrays are well conditioned and support substantially more angle pixels than physical antennas.

    \item \textit{Wideband angle-imaging framework with frequency-dependent channels:}
    We formulate a wideband angle-imaging model in which a dense angular response is reconstructed over a prescribed FoV. 
    The model accounts for frequency-dependent spatial steering, interprets each frequency-antenna pair as a virtual spatial sample, and shows that the nominal angular resolution is governed by the highest operating frequency rather than the center frequency. 
    We further incorporate frequency-dependent scattering responses through a finite basis expansion, yielding a unified framework for both flat and frequency-dependent angular responses.
\end{itemize}

\subsection{Notation and Organization}

Throughout this paper, uppercase boldface letters $\bm{X}$, lowercase boldface letters $\bm{x}$, and lowercase non-bold letters $x$ denote matrices, vectors, and scalars, respectively. 
Superscripts $(\cdot)^\T$ and $(\cdot)^\H$ denote the transpose and conjugate transpose. 
Calligraphic letters, e.g., $\mathcal{X}$, denote finite ordered sets, with elements arranged in ascending order. 
In particular, $\mathcal{I}(N)\triangleq\{1,\ldots,N\}$. 
For a matrix, $\lambda_i(\cdot)$ and $\sigma_i(\cdot)$ denote its $i$-th eigenvalue and singular value, respectively. 
The identity matrix of size $N$ is denoted by $\bm{I}_N$, and $j\triangleq\sqrt{-1}$.

The remainder of this paper is organized as follows. 
Section~\ref{sec: signal framework} presents the wideband angle-imaging signal model with frequency-dependent channels. 
Section~\ref{sec: properties} discusses the wideband virtual-aperture properties and the associated stability challenge. 
Section~\ref{sec: CC} introduces the proposed CC and its theoretical analysis. 
Section~\ref{sec: nonlinear array design} develops the CC-guided non-uniform sparse-array design. 
Section~\ref{sec: simulation} presents numerical results, and Section~\ref{sec: conclusion} concludes the paper.

\section{Wideband Angle Imaging Framework}
\label{sec: signal framework}
\subsection{Transmitted Signal Model}
\label{subsec: transmitted signal model}
Consider a wideband probing signal occupying the frequency interval $[f_L, f_H]$ with bandwidth $B = f_H - f_L$. The interval is sampled at $M_f$ uniformly spaced frequencies with spacing $\Delta_f = B/(M_f-1)$. Ordered from high to low frequency, the full frequency set is
\begin{equation}
    \mathcal{F}\triangleq\{f_{m_f}\}_{m_f\in\mathcal{I}(M_f)},\quad f_{m_f}=f_H-(m_f-1)\Delta_f.
    \label{eq: frequency set def}
\end{equation}

The signal with period $T_c$ is transmitted using one transmitting antenna, and the tone at frequency $f_{m_f}$ is 
\begin{equation}
    x(f_{m_f}, t) = e^{j2\pi f_{m_f} t },\quad t\in[0, T_c).
    \label{eq: trans signal}
\end{equation}

\subsection{Frequency-Dependent Wireless Channel}
\label{subsec: Frequency-Dependent Wireless Channel}
In narrowband systems, the channel response is commonly approximated as flat over the signal bandwidth. In contrast, different frequency components in wideband signals may experience different propagation responses. 

To quantify channel fluctuation over frequency,  we model the total phase excursion across the $M_f$ frequency samples as $\Delta\Phi \triangleq 2\pi r(M_f-1)$, where $r\in[0,1]$ is the frequency-variation ratio. We define the channel variation complexity as
\[
    N_b \triangleq \left\lceil r(M_f-1)\right\rceil +1,\quad N_b\in \mathbb Z_+.
\]
The case $N_b = 1$ corresponds to a flat channel response.

The frequency-dependent channel response is modeled by a finite-dimensional Fourier-series expansion \cite[eq.~(3)]{10447609},
\begin{equation}
    \gamma(f_{m_f}, \theta) 
    = \sum_{n_b=1}^{N_b} g(\theta, n_b)
    e^{-j2\pi (m_f-1) (n_b-1) / M_f},
    \label{eq: channel response}
\end{equation}
where $\gamma(f_{m_f},\theta)$ incorporates path loss, phase delay, reflection, scattering, and the target radar cross section. The channel coefficient $g(\theta, n_b)$ denotes the contribution associated with angle $\theta$ and frequency basis index $n_b$.

\subsection{Received Signal Model}
\label{subsec: received signal model}
The received signal consists of the superposition of reflections from all angles within the supported FoV \footnote{The definition of supported FoV is given in Section \ref{subsec: challenges}.} and across all frequency samples. This signal is observed by a receiving array with width $W$ and $M_a$ antennas. Let $p_{m_a}$ denote the position of the $m_a$-th antenna along the array axis, where $m_a\in\mathcal{I}(M_a)$ and $p_1=0$ is chosen as the reference position. We use the set $\mathcal{P}_a\triangleq \{p_{m_a}~|~ m_a\in\mathcal{I}(M_a)\}$ to denote the antenna array. 

Let $\Theta\subseteq[-\pi/2,\pi/2]$ denote the supported FoV in radians. Denoting $u\triangleq \sin\theta$ as the direction cosine, the supported FoV in the $u$-domain is
\[
\mathcal{U}\triangleq \{\sin\theta:\theta\in\Theta\}=[u_{\min},u_{\max})\subseteq[-1,1].
\]
For angle imaging, the angular response over the FoV $\mathcal U$ is discretized into uniformly spaced angle pixels, and we define the angular grid as
\begin{equation}
    \mathcal{G}_\theta
    \triangleq
    \{n_\theta\in\mathbb{Z}: u_{\min}\le n_\theta\Delta_u<u_{\max}\},
    \quad
    N_\theta\triangleq|\mathcal{G}_\theta|
    \label{eq: angle grid}
\end{equation}
where $\Delta_u$ denotes the angle spacing in the $u$-domain. In this work, we set $\Delta_u$ equal to the theoretical wideband angular resolution,\footnote{We provide a formal introduction of $u_{\mathrm{res}}$ in Section~\ref{subsec: opportunity}.} i.e., 
\[
    \Delta_u = u_{\mathrm{res}} = \frac{c}{Wf_H}.
\] 

After extracting the frequency component at $f_{m_f}$ through temporal Fourier processing, the noisy observation is written as
\begin{equation}
    \widetilde{\bm y}_{m_f}
    =
    \bm y_{m_f}+\bm w_{m_f},
    \qquad
    \bm y_{m_f}
    =
    \bm A_{\theta,m_f}\bm{\gamma}_{m_f},
    \label{eq: received signal 1}
\end{equation}
where $\bm w_{m_f}\in\mathbb C^{M_a\times 1}$ denotes the additive white Gaussian noise. The channel-response vector is defined as
\[
    \bm{\gamma}_{m_f}
    \triangleq
    [\gamma(f_{m_f},n_\theta)]_{n_\theta\in\mathcal G_\theta}^{\T}
    \in\mathbb C^{N_\theta\times 1},
\]
whose entries follow the frequency-dependent channel model in \eqref{eq: channel response}. The angle steering matrix is given by
\begin{equation}
    \bm A_{\theta,m_f}
    \triangleq
    \big[\bm a_{m_f}(n_\theta)\big]_{n_\theta\in\mathcal G_\theta}
    \in\mathbb C^{M_a\times N_\theta},
    \label{eq: angle steering matrix}
\end{equation}
with
\begin{equation}
    \bm a_{m_f}(n_\theta)
    \triangleq
    [e^{-j2\pi f_{m_f}p_{m_a}n_\theta/(Wf_H)}]_{m_a\in\mathcal I(M_a)}^{\T}
    \in\mathbb C^{M_a\times 1}.
    \label{eq: steering vector}
\end{equation}

According to the channel expansion in \eqref{eq: channel response}, the channel-response vector can be further written as
\begin{equation}
    \bm{\gamma}_{m_f}
    =
    \bm G\bm b_{m_f}^{\T},
    \label{eq: gamma vector basis}
\end{equation}
where $\bm G\in\mathbb C^{N_\theta\times N_b}$ collects the channel coefficients, and
\[
    \bm b_{m_f}
    \triangleq
    [e^{-j2\pi (m_f-1)(n_b-1)/M_f}]_{n_b\in\mathcal{I}(N_b)}
    \in\mathbb C^{1\times N_b}.
\]
Substituting \eqref{eq: gamma vector basis} into \eqref{eq: received signal 1} gives
\begin{equation}
    \bm y_{m_f}
    =
    \bm A_{\theta,m_f}\bm G\bm b_{m_f}^{\T}.
    \label{eq: received signal 2}
\end{equation}

Stacking $\bm y_{m_f}$ from all frequency samples, we obtain
\begin{equation}
    \bm y \triangleq \begin{bmatrix}
        \bm y_1^\T& \cdots & 
        \bm y_{M_f}^\T
    \end{bmatrix}^\T = \bm A^{\fd}\operatorname{vec}(\bm G) \in\mathbb{C}^{M_fM_a\times 1},
    \label{eq: received signal 3}
\end{equation}
where
\begin{equation}
    \bm A^{\fd} \triangleq
    \begin{bmatrix}
        \bm b_1\otimes \bm A_{\theta,1}\\
        \vdots\\
        \bm b_{M_f}\otimes \bm A_{\theta,M_f}
    \end{bmatrix}
    \in\mathbb{C}^{M_fM_a\times N_bN_\theta}
    \label{eq: system matrix_select}
\end{equation}
is the system matrix under a frequency-dependent channel. Specifically, when $N_b = 1$, it reduces to the flat-channel system
\begin{equation}
    \bm A^{\fl}\triangleq \begin{bmatrix}
        \bm A_{\theta,1}^\T & \cdots & 
        \bm A_{\theta,M_f}^\T
    \end{bmatrix}^\T \in\mathbb{C}^{M_fM_a\times N_\theta}
    \label{eq: system matrix_flat}.
\end{equation}

\subsection{Inverse problem for imaging}

Angle imaging is formulated as an inverse problem of recovering $\bm G$ from $\bm y$ through the linear system \eqref{eq: received signal 3}. The coefficient matrix can then be estimated as
\begin{equation}
\label{eq: inverse problem}
    \bm G^\star = \arg \min_{\bm G}\|\widetilde{\bm y} - \bm A^\fd \operatorname{vec}(\bm G)\|_2^2,
\end{equation}
where $\widetilde{\bm y}\triangleq [\widetilde{\bm y}_{1}^\T \cdots \widetilde{\bm y}_{M_f}^\T]^\T$ represents the noisy measurements.
The angular response $\gamma(f_{m_f},n_\theta)$ can be reconstructed from $\bm G^\star$ using \eqref{eq: channel response}. 

\section{Opportunities and Challenges of Wideband Angle Imaging}
\label{sec: properties}

From \eqref{eq: steering vector}, each physical antenna-frequency sample pair can be interpreted as a virtual spatial sample. 
Hence, wideband angle imaging can be equivalently viewed as imaging over a frequency-induced virtual array. 
This provides the key opportunity of wideband signaling: the additional virtual samples enlarge the effective spatial sampling set, allowing a sparse physical array to support more angle pixels and mitigate the spatial aliasing that would occur in a sparse narrowband array.

However, these virtual samples are generally non-uniformly distributed, which may create large gaps over the virtual aperture. Such irregular spatial sampling can make the system matrix ill-conditioned and lead to unstable recovery.

This section first formalizes the virtual-array interpretation and then
discusses the resulting opportunities and challenges in wideband angle imaging.

\subsection{Virtual-Array Interpretation}

\begin{figure}
    \centering
    \includegraphics[width=0.9\linewidth]{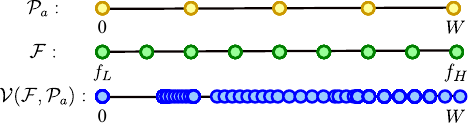}
    \caption{Illustration of the virtual-array geometry. The formed virtual array may contain large gaps between adjacent virtual elements, leading to a poorly conditioned imaging system and unstable recovery performance.}
    \label{fig: virtual array geometry}
\end{figure}

\begin{definition}[Virtual Array Definition]
Given a frequency sample set $\mathcal{F}$ and a physical linear array $\mathcal{P}_a$, define the virtual array as the collection of virtual elements
\begin{equation}
\begin{aligned}
    \mathcal{V}(\mathcal{F}, \mathcal{P}_a) \triangleq
    & \left\{\frac{fp}{f_H}~|~ f\in \mathcal{F}, p\in \mathcal{P}_a\right\}.
\end{aligned}
    \label{eq: virtual array def}
\end{equation}
Repeated values are retained as separate virtual samples. 
\end{definition}
The virtual array has the same width as the physical array, since the largest virtual location is attained at $f=f_H$. 
However, its sampling locations are induced jointly by frequency and spatial samples, and are therefore generally much denser and non-uniform even when both the physical antennas and frequency samples are uniformly spaced. 
Fig.~\ref{fig: virtual array geometry} illustrates this construction.

\subsection{Opportunities}
\label{subsec: opportunity}

\paragraph{Improved angular resolution}

For a narrowband array with width $W$ and carrier frequency $f_c$, it is well-known that the angular resolution in the $u$-domain is approximately
\begin{equation}
    u_{\mathrm{res,narrow}}
    \triangleq
    \min |\sin\theta_1-\sin\theta_2|
    \approx 
    \frac{2}{M_a} = \frac{c}{f_c W},
\end{equation}
where the antennas are uniformly distributed with spacing $c/(2f_c)$.

In wideband angle imaging, based on the Fourier uncertainty principle \cite{gabor1946theory, proakis2013digital}, the minimum resolvable separation in the angle domain is inversely proportional to the effective window width along the conjugate variable, which here is governed by $f_H W / c$.
Accordingly, the achievable angular resolution is
\begin{equation}
    u_{\mathrm{res}}
    \triangleq
    \frac{c}{f_H W}.
    \label{eq: wideband angular resolution}
\end{equation}
Hence, compared with the narrowband case, wideband signaling improves the theoretical angular resolution by replacing $f_c$ with the larger frequency $f_H$.

\paragraph{Spatial aliasing mitigation}
Another key implication of wideband signaling is that it relaxes the classical half-wavelength spacing constraint, as frequency diversity helps distinguish aliased spatial responses across frequencies. As a result, wideband angle imaging can operate with much larger inter-element spacings, which is the key enabler for the sparse array design developed later.

\begin{figure}[t]
    \centering
    \includegraphics[width=0.9\linewidth]{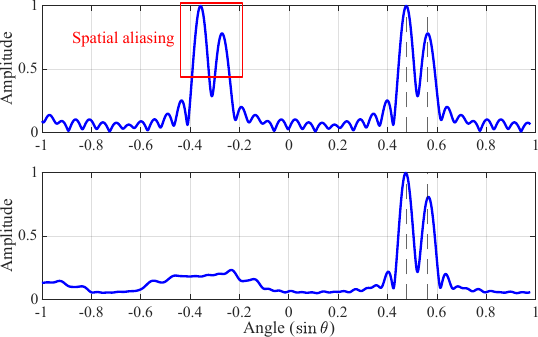}
    \caption{Angle-response comparison between the narrowband and wideband cases for a uniform linear array (ULA) with inter-element spacing $d=1.2c/f_c$. The top and bottom panels correspond to the narrowband and wideband cases, respectively.}
    \label{fig: Spatial aliasing}
\end{figure}

The aliasing mitigation of wideband signaling is illustrated in Fig.~\ref{fig: Spatial aliasing}. Under the same sparse array configuration, the narrowband response exhibits prominent aliasing peaks, whereas the wideband response remains free of such ambiguities.

\subsection{Challenges}
\label{subsec: challenges}

The above opportunities do not automatically guarantee stable image recovery.
Although wideband signaling introduces more virtual elements, these elements are generally irregularly distributed over the virtual aperture, as illustrated in Fig.~\ref{fig: virtual array geometry}.
In particular, large gaps may appear between adjacent virtual elements, even when the total number of virtual elements is large.
Such irregular spatial-frequency sampling can make the columns of the system matrix highly correlated, leading to an ill-conditioned system matrix.
As a result, the image vector $\bm G$ in \eqref{eq: inverse problem} cannot be stably recovered.

Therefore, the stable imaging capability should be characterized by the number of angle pixels that can be supported with a well-conditioned system matrix. 
To this end, we consider a contiguous set of angle pixels with spacing $u_{\mathrm{res}}$ and define the corresponding supported FoV.

\begin{definition}[Supported FoV]
Given a virtual array, suppose that the associated system matrix stably supports $N_\theta$ contiguous angle pixels. Here, ``stable'' means that the associated system matrix admits a controlled upper bound on its spectral condition number.

Then, the supported FoV is defined as the corresponding contiguous interval in the $u$-domain, whose width is approximately $N_\theta u_{\mathrm{res}}$.
\label{def: supported FoV}
\end{definition}
The system matrix is then constructed over the angle pixels within this supported FoV. Once $N_\theta$ is specified, the sensing setup can be configured for the associated angular sector of interest. If the supported FoV covers the entire angular domain, the system is said to support full-FoV angle imaging.

Accordingly, the central challenge in wideband angle imaging is to determine the maximum $N_\theta$ that can be stably recovered, in the sense that the associated system matrix admits a controlled condition-number bound.
This motivates the CC developed in the next section.

\section{Coverage Criterion for Stable Wideband Angle Imaging}
\label{sec: CC}

In this section, we introduce the CC, an singular-value-decomposition (SVD)-free and geometry-driven rule that characterizes how many contiguous angle pixels can be stably supported by a given wideband sensing configuration.
Instead of directly evaluating the condition number of the system matrix for each candidate angular grid, the CC uses the maximum gap of the effective virtual array as a tractable geometric surrogate. 

We further show that, under the CC, the associated system matrices admit explicit condition-number upper bounds, thereby providing a stability guarantee for wideband angle imaging.


\subsection{Coverage Criterion}
\label{subsec: CC}
The intuition behind the CC is to determine how many angle pixels can be supported before the virtual array becomes under-sampled \cite{patwari2021sparse}.
Specifically, for a virtual array with width $W$, supporting $N_\theta$ angle pixels requires a nominal virtual sampling interval of $W/N_\theta$.
Hence, the gaps between adjacent virtual elements should not exceed $W/N_\theta$; otherwise, part of the virtual aperture is insufficiently sampled.

To expose the system structure underlying the CC, we begin by reformulating the system matrix $\bm A^\fd$. Since $\bm b_{m_f}\in\mathbb{C}^{1\times N_b}$ lies in an $N_b$-dimensional complex vector space, it can be expanded over an orthogonal basis of $\mathbb{C}^{N_b}$. Here, we adopt the rows of the $N_b$-point discrete Fourier transform (DFT) matrix as the orthogonal basis. These rows are denoted by $\{\bm e_i\}_{i=1}^{N_b}$ with entries $[\bm e_i]_m = e^{-j2\pi (i-1)(m-1)/N_b}$.
Accordingly, $\bm b_{m_f}$ admits the expansion
\begin{equation}
\label{eq: expansion of b}
    \bm b_{m_f}=\sum_{i=1}^{N_b}\beta_{i,m_f}\bm e_i = \bm \beta_{m_f}\bm E,
\end{equation}
where $\bm E$ collects the DFT rows and 
\begin{equation}
\label{eq: beta computation}
    \beta_{i,m_f}=\frac{1}{N_b}\bm b_{m_f}\bm e_i^{\H} = \frac{1}{N_b}e^{j \varphi_{i, m_f}}
    \frac{\sin\!\left(\pi N_b\Delta_{i,m_f}\right)}
         {\sin\!\left(\pi \Delta_{i,m_f}\right)}.
\end{equation}
Here, 
\[
    \Delta_{i,m_f}\triangleq \frac{i-1}{N_b}-\frac{m_f-1}{M_f}, \quad
    \varphi_{i,m_f} = -\pi (N_b-1)\Delta_{i, m_f},
\]
and 
\begin{align*}
    \bm \beta_{m_f}\triangleq \bigl[\beta_{1,m_f} \cdots \beta_{N_b,m_f} \bigr]\in\mathbb{C}^{1\times N_b}.
\end{align*}
From \eqref{eq: beta computation}, each coefficient sequence $\{\beta_{i, m_f}\}_{m_f\in\mathcal{I}(M_f)}$ follows a Dirichlet-kernel profile over the frequency samples, and sequences associated with different $i$ are mutually orthogonal, as illustrated in Fig.~\ref{fig: orthogonality of beta}.

Substituting \eqref{eq: expansion of b} into $\bm A^{\fd}$ yields
\begin{equation}
\label{eq: system matrix 3}
    \bm A^{\fd}=
    \underbrace{\begin{bmatrix}
        \bm \beta_{1}\otimes \bm A_{\theta, 1}\\
        \vdots\\
        \bm \beta_{M_f}\otimes \bm A_{\theta, M_f}
    \end{bmatrix}}_{\widetilde{\bm A}^{\fd}}(\bm E\otimes\bm I_{N_\theta}).
\end{equation}
Since $\bm \beta_{m_f}\otimes \bm A_{\theta, m_f} = \bigl[\beta_{1,m_f}\bm A_{\theta, m_f} \cdots \beta_{N_b,m_f}\bm A_{\theta, m_f} \bigr]$, $\widetilde{\bm A}^{\fd}$ can be represented by
\begin{equation}
    \widetilde{\bm A}^{\fd} = \bigl[\bm D_1\bm A^{\fl} \cdots \bm D_{N_b}\bm A^{\fl} \bigr]\in\mathbb{C}^{M_aM_f\times N_bN_\theta}
    \label{eq: equivalent system matrix}
\end{equation}
where 
\[
    \bm D_i \triangleq \operatorname{diag}(\beta_{i,1} \cdots \beta_{i,M_f})\otimes \bm I_{M_a}\in\mathbb{C}^{M_fM_a\times M_fM_a}.
\]

\eqref{eq: equivalent system matrix} shows that the imaging model can be viewed as the concatenation of $N_b$ weighted flat-channel components. 
For each $i\in\mathcal{I}(N_b)$, $\bm D_i\bm A^{\fl}$ represents a frequency-weighted version of the flat-channel steering matrix associated with the virtual array $\mathcal{V}(\mathcal{F},\mathcal{P}_a)$. 
Specifically, the rows corresponding to the virtual elements induced by frequency $f_{m_f}$ are weighted by the coefficient $\beta_{i,m_f}$. 
Hence, the magnitude $|\beta_{i,m_f}|$ quantifies how strongly the virtual elements generated by $f_{m_f}$ contribute to the $i$-th component.

Based on the structure of \eqref{eq: equivalent system matrix}, we now introduce the CC.

\begin{definition}[Coverage Criterion]
\label{def: CC}
    Consider a frequency set $\mathcal F$, physical array $\mathcal P_a$, array width $W$ and channel variation complexity $N_b$ for angle imaging. Let $\epsilon\in(0,1)$ be a prescribed threshold. 
    For each $i\in\mathcal I(N_b)$, define the active frequency set as
    \begin{equation}
    \label{eq: active freq set def}
    \mathcal F_i(\epsilon)
    \triangleq
    \left\{
    f_{m_f}\in\mathcal F
    \;\middle|\;
    |\beta_{i,m_f}|>\epsilon
    \right\}.
    \end{equation}
    The corresponding effective virtual array is defined by
    \begin{equation}
    \label{eq: effective virtual array def}
    \mathcal V_i^\epsilon
    \triangleq
    \left\{
    \frac{fp}{f_H}
    \;\middle|\;
    f\in\mathcal F_i(\epsilon),\;
    p\in\mathcal P_a
    \right\}.
    \end{equation}
    For $N_b=1$, $\mathcal F_1(\epsilon)=\mathcal F$, and $\mathcal V_1^\epsilon$ reduces to the virtual array in \eqref{eq: virtual array def}.
    
    For $x,v\in[0,W)$, define the wrap-around distance
    \begin{equation}
    \label{eq: wrap_dist_def}
    \operatorname{dist}(x,v)
    \triangleq
    \min\{|x-v|,\;W-|x-v|\}.
    \end{equation}
    
    Given a coverage length $C\in(0,W]$, the covered region induced by 
    $\mathcal V_i^\epsilon$ is
    \begin{equation}
    \label{eq: S_def}
    \mathcal S_i(C) \triangleq \bigcup_{v\in\mathcal V_i^\epsilon} \left\{ x\in[0,W) \;\middle|\; \operatorname{dist}(x,v)\le \frac{C}{2} \right\}.
    \end{equation}
    In other words, each virtual element covers a circular interval of length $C$, centered at its location.
    
    The minimum coverage length required to cover the whole virtual aperture is defined as 
    \begin{equation} 
    \label{eq: min coverage length} 
    C_i^\star \triangleq \inf\left\{ C\in(0,W] \;\middle|\; \mathcal S_i(C)=[0,W) \right\}. 
    \end{equation}
    
    Equivalently, $C_i^\star$ is given by the maximum wrap-around inter-element spacing of $\mathcal V_i^\epsilon$. Denoting this spacing by $d_{\max,i}$, we have $C_i^\star = d_{\max, i}$.
    
    Accordingly, the CC estimates the number of angle pixels that can be stably supported as 
    \begin{equation} 
    \label{eq: actual angle pixels minimum} 
    N_\theta^{\cc} \triangleq \min_{i\in\mathcal I(N_b)} \left\{ \min\left\{ N_{\theta,\max},\; \left\lfloor \frac{W}{d_{\max,i}} \right\rfloor \right\} \right\}, 
    \end{equation}
    where $N_{\theta,\max}\triangleq \lfloor 2Wf_H/c \rfloor$ denotes the number of angle pixels over the full FoV.
\end{definition}

Specifically, \eqref{eq: min coverage length} specifies the CC coverage condition for each effective virtual array. Together with \eqref{eq: actual angle pixels minimum}, the CC implies that supporting $N_\theta$ angle pixels requires
\begin{equation}
\label{eq: coverage requirement}
C_i^\star=d_{\max,i}\le \frac{W}{N_\theta},
\qquad \forall i\in\mathcal I(N_b).
\end{equation}
This is consistent with the design intuition of the CC, where the maximum inter-element spacing in each $\mathcal{V}_i^\epsilon$ cannot exceed $W/N_\theta$.

The threshold $\epsilon$ defined in Definition~\ref{def: CC} controls the size of each effective virtual array and, consequently, the supported $N_\theta^\cc$. In particular, a larger $\epsilon$ leads to a smaller active frequency set $\mathcal F_i(\epsilon)$ and hence a smaller effective virtual array $\mathcal V_i^\epsilon$, which yields a smaller value of $N_{\theta}^\cc$. Conversely, a smaller $\epsilon$ retains more frequency samples and may allow a larger $N_{\theta}^\cc$. 

\begin{figure}
    \centering
    \includegraphics[width=\linewidth]{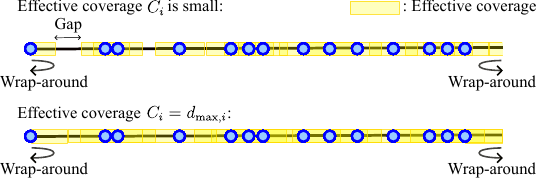}
    \caption {Illustration of the CC for the $i$-th effective virtual array $\mathcal V_i^\epsilon$. When $C_i < d_{\max, i}$, gaps remain in the covered region. The full coverage condition \eqref{eq: min coverage length} is achieved only when $C_i \geq d_{\max, i}$. This condition must hold for all $i\in\mathcal I(N_b)$.}
    \label{fig: CC illustration}
\end{figure}

Fig.~\ref{fig: CC illustration} illustrates the CC for any effective virtual array $\mathcal V_i^\epsilon$.

The CC-designed system matrix has $M_aM_f$ rows and $N_b N_\theta^{\cc}$ columns. 
The following subsections analyze its conditioning.

\subsection{Conditioning Analysis under Flat Channels}
\label{subsec: cond analysis}

We now provide a condition-number interpretation of the CC under flat channels.
For a full-column-rank system matrix $\bm A^\fl$, we use the spectral condition number $\kappa(\bm A)\triangleq \sigma_{\max}(\bm A)/\sigma_{\min}(\bm A)$ to quantify the numerical stability of the corresponding problem.
In this paper, a system matrix is said to be stable if its spectral condition number admits a controlled upper bound.

Consider the virtual array in \eqref{eq: virtual array def}, written as
\begin{equation}
    \mathcal V(\mathcal F,\mathcal P_a)
    =
    \{v_{m_v}\}_{m_v\in\mathcal I(M_v)},\quad M_v\triangleq M_fM_a,
\label{eq: virtual array def 2}
\end{equation}
where
\[
    0\le v_1\le v_2\le \cdots\le v_{M_v}\le W .
\]
For a flat channel, the received signal presented in \eqref{eq: received signal 3} is simplified to 
\begin{equation}
\label{eq: flat vector model}
    \bm y = \bm A^\fl \bm \gamma,
\end{equation}
where the entry in $\bm A^\fl$ can be represented using the virtual array, i.e., 
\begin{equation}
    [\bm A^\fl]_{m_v, n_\theta} = e^{-j2\pi v_{m_v}n_\theta/W},\quad m_v\in\mathcal{I}(M_v),~ n_\theta\in\mathcal{G}_\theta. 
\end{equation}

To relate the non-uniform virtual-array geometry to a deterministic conditioning bound, we introduce the Voronoi weights of the virtual elements. Specifically, the circular Voronoi interval for $v_{m_v}$ is defined by
\[
    I_{m_v}
    \triangleq
    \left[
    \frac{v_{m_v-1}+v_{m_v}}{2},
    \frac{v_{m_v}+v_{m_v+1}}{2}
    \right), \quad m_v\in\mathcal{I}(M_v)
\]
with $v_0 \triangleq v_{M_v}-W$ and $v_{M_v+1}\triangleq v_1+W$. Then, the corresponding Voronoi weights and the corresponding weighted matrix are defined by
\begin{equation}
\label{eq: Voronoi weights}
    w_{m_v} \triangleq |I_{m_v}|, \quad 
    \bm W_v
    \triangleq
    \operatorname{diag}(w_1,\ldots,w_{M_v}).
\end{equation}
The Voronoi-weighted flat-channel matrix is defined as
\begin{equation}
    \widetilde{\bm A}^{\fl}
    \triangleq
    \bm W_v^{1/2}\bm A^{\fl}.
    \label{eq: weighted flat channel matrix}
\end{equation}
Accordingly, the original observation model \eqref{eq: flat vector model} can be
equivalently row-weighted as
\begin{equation}
    \widetilde{\bm y} 
    \triangleq
    \bm W_v^{1/2}\bm y 
    =
    \widetilde{\bm A}^{\fl}\bm\gamma.
    \label{eq: weighted flat vector model}
\end{equation}
This weighting does not require additional measurements; it is a deterministic row scaling of the original system induced by the virtual-array geometry.

\begin{theorem}[Upper bound on $\kappa(\widetilde{\bm A}^\fl)$]
\label{theorem: cc weighted conditioning flat}
Consider a flat-channel system matrix $\bm A^{\fl}\in\mathbb C^{M_v\times N_\theta^\cc}$ generated by the virtual array in \eqref{eq: virtual array def 2}. Assume that the angular grid $\mathcal G_\theta$ contains $N_\theta^\cc$ consecutive integer indices, where $N_\theta^\cc$ is selected according to the CC. Let $d_{\max}$ denote the maximum circular inter-element gap of the virtual array.

Generate the Voronoi-weighted flat-channel matrix
$\widetilde{\bm A}^{\fl}$ by \eqref{eq: weighted flat channel matrix}. Then, its condition number satisfies
\begin{equation}
    \kappa\!\left(\widetilde{\bm A}^{\fl}\right)
    \le
    2\frac{W}{d_{\max}}-1 .
    \label{eq: cc weighted condition number bound}
\end{equation}
\end{theorem}

Proof is shown in Appendix \ref{Proof: cc weighted conditioning flat}.
Theorem~\ref{theorem: cc weighted conditioning flat} gives a deterministic condition-number interpretation of the CC. It shows that the maximum virtual-array gap not only determine the value of $N_\theta^\cc$, but also yields
an explicit upper bound on $\kappa(\widetilde{\bm A}^\fl)$.

Since the weighted system is obtained from the original observations by applying the scaling matrix $\bm W_v^{1/2}$, the angular response $\bm \gamma$ to be recovered is exactly the same as the original system. In this sense, the inverse problem for angle imaging can be equivalently processed using the weighted system, 
\begin{equation}
    \bm\gamma^\star
    =
    \arg\min_{\bm\gamma}
    \left\|
    \bm W_v^{1/2}\bm y-\widetilde{\bm A}^{\fl}\bm\gamma
    \right\|_2^2,
    \label{eq: voronoi weighted ls}
\end{equation}
whose system matrix is precisely controlled by
Theorem~\ref{theorem: cc weighted conditioning flat}. 

Therefore, in the flat-channel case, the CC provides an SVD-free route to characterize the supportable angular-grid size, while the stability of the corresponding imaging system is quantified through an explicit condition-number upper bound for its associated weighted matrix.

\subsection{Conditioning analysis under frequency-dependent channels}
\label{subsec: cond analysis for A_fd}

We now extend the conditioning analysis to frequency-dependent channels. Recall the frequency-dependent system matrix in \eqref{eq: system matrix 3}. Following the flat-channel analysis, we define the Voronoi-weighted system matrix for non-flat case as
\begin{equation} 
    \bm A_{\mathrm w}^{\fd} \triangleq \bm W_v^{1/2} \bm A^{\fd} = \bm W_v^{1/2}\widetilde{\bm A}^\fd(\bm E\otimes \bm I_{N_\theta}).
    \label{eq: weighted fd matrix} 
\end{equation}
Since $\bm W_v^{1/2}$ and $\bm D_i$ defined in $\widetilde{\bm A}^\fd$ are both diagonal matrices, we have 
\begin{equation} 
    \begin{aligned} 
        \bm A_{\mathrm w}^{\fd} 
        &= \left[ \bm D_1\bm W_v^{1/2}\bm A^{\fl}, \ldots, \bm D_{N_b}\bm W_v^{1/2}\bm A^{\fl} \right](\bm E\otimes \bm I_{N_\theta}) \\
        &= \widetilde{\bm A}^\fd_{\mathrm w}(\bm E\otimes \bm I_{N_\theta}),
    \end{aligned} 
\label{eq: weighted fd block form} 
\end{equation}
where 
\begin{equation}
    \widetilde{\bm A}^\fd_{\mathrm w} \triangleq \left[ \bm D_1\widetilde{\bm A}^{\fl}, \ldots, \bm D_{N_b}\widetilde{\bm A}^{\fl} \right].
    \label{eq: weighted fd block without unitary}
\end{equation}

Similar to the flat-channel case, the weighted system does not change the angular-response coefficients to be recovered. Therefore, the following analysis studies the stability of $\bm A_{\mathrm w}^{\fd}$.

The analysis of $\kappa(\bm A_{\mathrm w}^{\fd})$ is carried out in two steps. We first bound each basis-wise matrix $\bm D_i\widetilde{\bm A}^\fl$, characterizing its self-correlation energy. We then analyze their concatenation in $\widetilde{\bm A}_{\mathrm w}^{\fd}$ by relating the cross-correlation energy among different basis-wise components to the resulting condition-number bound.

\begin{lemma}[Conditioning of $\bm D_i\widetilde{\bm A}^\fl$]
    \label{lem: cond of BiA}
    Consider a CC-designed system matrix expressed as in \eqref{eq: system matrix 3} for recovering $N_{\theta}^\cc$ angle pixels, with threshold $\epsilon\in(0,1)$.
    
    For $i\in\mathcal{I}(N_b)$, let $\widetilde{\bm A}_i$ be the row-submatrix of $\widetilde{\bm A}^{\fl}$ associated with the active frequency set $\mathcal F_i(\epsilon)$. 
     
    Then, each $\widetilde{\bm A}_i$ is a full-column rank matrix, and
    \begin{equation}
    \begin{aligned}
        \sigma^2_{\min}(\bm D_i\widetilde{\bm A}^{\fl}) &\geq \epsilon^2\sigma^2_{\min}(\widetilde{\bm A}_i),\\
        \sigma_{\max}^2(\bm D_i\widetilde{\bm A}^{\fl}) &\le \sigma_{\max}^2(\widetilde{\bm A}_i) + \epsilon^2 \left( \sigma_{\max}^2(\widetilde{\bm A}^\fl) - \sigma_{\min}^2(\widetilde{\bm A}_i) \right).    
    \end{aligned}
    \label{eq: bound of DiA}
    \end{equation}
\end{lemma}

Proof is provided in Appendix \ref{proof: cond of DiA}. The bound in \eqref{eq: bound of DiA} depends on the singular values of $\widetilde{\bm A}_i$, whose condition number admits an upper bound by Theorem~\ref{theorem: cc weighted conditioning flat}. Moreover, compared with $\widetilde{\bm A}_i$, $\widetilde{\bm A}^{\fl}$ retains all virtual elements for the same $N_\theta^\cc$, hence, it also admits controlled condition-number bounds. Combining these results with \eqref{eq: bound of DiA} yields controlled singular-value bounds for each basis-wise matrix $\bm D_i\widetilde{\bm A}^\fl$. 

The $\epsilon$-dependent lower bound in $\sigma_{\min}(\bm D_i\widetilde{\bm A}^{\fl})$ is consistent with the role of $\epsilon$ in the CC.
For a fixed number of measurements, a smaller $\epsilon$ may support a larger $N_\theta^\cc$, but this requires the same measurements to resolve more angle pixels and therefore weakens the spectral floor of each basis-wise block. A larger $\epsilon$ leads to a more conservative
$N_\theta^\cc$, but provides a stronger spectral floor and hence better conditioning.

\begin{theorem}[Upper bound on $\kappa(\bm A_{\mathrm w}^{\fd})$] 
\label{theorem: bound of A_fd}
Consider the weighted frequency-dependent system matrix $\bm A_{\mathrm w}^{\fd}$ in \eqref{eq: weighted fd block form} and $\widetilde{\bm A}_{\mathrm{w}}^{\fd}$ in \eqref{eq: weighted fd block without unitary}. Define the diagonal-block part of $( \widetilde{\bm A}^\fd_{\mathrm w})^\H \widetilde{\bm A}^\fd_{\mathrm w}$ as \begin{equation} 
    \begin{aligned} 
        \bm Q_D \triangleq \operatorname{blkdiag}\bigl(
        (\bm D_1\widetilde{\bm A}^{\fl})^\H \bm D_1\widetilde{\bm A}^{\fl}, \ldots, (\bm D_{N_b}\widetilde{\bm A}^{\fl})^\H \bm D_{N_b}\widetilde{\bm A}^{\fl} \bigr). 
    \end{aligned} 
\end{equation}
The off-diagonal-block part is defined as 
\begin{equation} 
    \bm Q_R \triangleq (\widetilde{\bm A}^\fd_{\mathrm w})^\H\widetilde{\bm A}^\fd_{\mathrm w} - \bm Q_D . 
\end{equation} 
Denote its spectral norm by $\delta\triangleq\|\bm Q_R\|_2$. If \begin{equation} \label{eq: sufficient condition for cond of A_fd} \delta < \min_{i\in\mathcal I(N_b)} \sigma_{\min}^2(\bm D_i\widetilde{\bm A}^{\fl}), \end{equation} then the condition number of $\bm A_{\mathrm w}^{\fd}$ is upper bounded by \begin{equation} 
    \kappa(\bm A_{\mathrm w}^{\fd}) \le \sqrt{ \frac{ \max_{i\in\mathcal I(N_b)} \sigma_{\max}^2(\bm D_i\widetilde{\bm A}^{\fl})+\delta }{\min_{i\in\mathcal I(N_b)} \sigma_{\min}^2(\bm D_i\widetilde{\bm A}^{\fl})-\delta } }. 
    \label{eq: bound of A_fd} 
    \end{equation} 
\end{theorem}

Proof is provided in Appendix~\ref{proof: bound of A_fd}. Theorem~\ref{theorem: bound of A_fd} shows that the stability of the weighted frequency-dependent system is determined by the spectral floor of the diagonal blocks and the leakage level $\delta$ of the off-diagonal part.

Although the sufficient condition \eqref{eq: sufficient condition for cond of A_fd} is not available in closed form, it is natural in the proposed design regime. For any $i,j\in\mathcal I(N_b)$, the $(i,j)$-th block of $(\widetilde{\bm A}_{\mathrm w}^{\fd})^\H\widetilde{\bm A}_{\mathrm w}^{\fd}$ is
\begin{equation} 
    (\bm D_i\widetilde{\bm A}^{\fl})^\H (\bm D_j\widetilde{\bm A}^{\fl}) = \sum_{m_f=1}^{M_f} \beta_{i,m_f}^\ast\beta_{j,m_f} \widetilde{\bm A}_{\theta,m_f}^\H \widetilde{\bm A}_{\theta,m_f}, 
    \label{eq: weighted Gij_block} 
\end{equation} 
where $\widetilde{\bm A}_{\theta,m_f}$ denotes the weighted angular steering matrix associated with the $m_f$-th frequency sample.

\begin{figure}[t]
\centering
\subfloat[]{%
    \includegraphics[width=0.49\columnwidth]{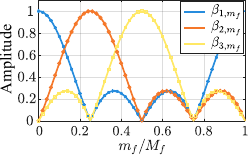}%
    \label{fig: orthogonality of beta}%
}
\hfill
\subfloat[]{%
    \includegraphics[width=0.49\columnwidth]{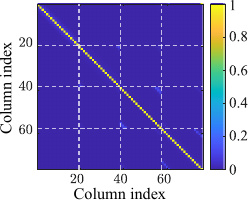}%
    \label{fig: Gram matrix of A_fd}%
}
\caption{Illustrations supporting Theorem~\ref{theorem: bound of A_fd}. 
(a) Representative coefficient profiles $\{\beta_{i,m_f}\}_{m_f\in\mathcal I(M_f)}$ for different basis indices, showing their orthogonality across frequency samples. 
(b) Representative Gram of a CC-designed system matrix. White dashed lines indicate the block boundaries of $(\bm D_i\widetilde{\bm A}^\fl)^\H(\bm D_j\widetilde{\bm A}^\fl)$.}
\label{fig:overall}
\end{figure}

If $i=j$, \eqref{eq: weighted Gij_block} corresponds to a diagonal block of $(\widetilde{\bm A}_{\mathrm w}^{\fd})^\H\widetilde{\bm A}_{\mathrm w}^{\fd}$. In this case, the coefficients reduce to $|\beta_{i,m_f}|^2$, leading to a positively weighted accumulation of per-frequency Gram matrices. In contrast, if $i\neq j$, the block is off-diagonal, and the coefficients $\beta_{i,m_f}^\ast\beta_{j,m_f}$ oscillate across frequency due to the orthogonality among different basis sequences, as illustrated in Fig.~\ref{fig: orthogonality of beta}.

Hence, the diagonal blocks collect the dominant energy through constructive accumulation, whereas the off-diagonal blocks are cancellation-dominated. 
As a result, the Gram matrix is expected to be approximately block diagonally dominant, so that the leakage level $\delta$ remains moderate relative to the diagonal-block spectral floor. A representative example is shown in Fig.~\ref{fig: Gram matrix of A_fd}, where the dominant energy is concentrated in the diagonal blocks and the off-diagonal coupling is weak.

Therefore, Theorem~\ref{theorem: bound of A_fd} gives explicit condition-number control whenever the inter-basis leakage $\delta$ remains below the diagonal-block spectral floor. Stable angle imaging is thus governed by CC-controlled basis-wise conditioning and residual inter-basis coupling.

\section{Non-Uniform Array Design}
\label{sec: nonlinear array design}

In this section, we develop a non-uniform array design that
allows all effective virtual arrays to satisfy the CC over a
common full-FoV angular grid. We first determine the finest
common angle spacing supported by all effective sub-bands.
We then derive the corresponding physical-antenna placement.

\begin{proposition}[Sparse Array Design]
For a channel with variation complexity $N_b$ over bandwidth $B$ (with highest frequency $f_H$), define the relative frequency ratio
\begin{equation}
    \alpha(N_b) \triangleq \frac{f_H}{B}\times N_b.
\label{eq: relative frequency ratio}
\end{equation}

To ensure full-FoV angle imaging with a budget of $M_a$ physical antennas, the $m_a\in\mathcal{I}(M_a)$-th antenna position is given by
\begin{equation} 
    p_{m_a} = \frac{c}{2 f_H}\times\left[\alpha(N_b)\Bigl(\frac{\alpha(N_b)}{ \alpha(N_b) - 1} \Bigr)^{m_a-1} - \alpha(N_b)\right].
\label{eq: non-uniform antenna location_fd} 
\end{equation}
The resulting physical array width is
\begin{equation} 
\begin{aligned} 
    W =& p_{M_a}-p_1\\ 
    =& \frac{c}{2 f_H}\times\left[\alpha(N_b)\Bigl(\frac{\alpha(N_b)}{ \alpha(N_b) - 1} \Bigr)^{M_a-1} - \alpha(N_b)\right] . 
\end{aligned} 
\label{eq: array width design}
\end{equation}
This array supports full-FoV angle imaging with angle spacing $\Delta_u = c/(Wf_{\star})$ with $f_{\star} \triangleq f_L + B/N_b$.
\end{proposition}
The design strategy is as follows.

To clearly illustrate the design, we select $\epsilon$ at the
intersection of two adjacent coefficient profiles
$\{\beta_{i,m_f}\}_{m_f\in\mathcal I(M_f)}$. Under this
choice, the active frequency sets form $N_b$ contiguous
sub-bands of equal width $B_{\mathrm{eff}}\triangleq B/N_b$.
For $i\in\mathcal I(N_b)$, the lowest and highest frequencies
of $\mathcal F_i(\epsilon)$ are
\begin{equation}
    f_L^{(i)}
    =
    f_H-iB_{\mathrm{eff}},
    \qquad
    f_H^{(i)}
    =
    f_H-(i-1)B_{\mathrm{eff}},
    \label{eq: effective subband edges}
\end{equation}
respectively.\footnote{These contiguous sub-bands can be
realized by using a shifted DFT basis in
\eqref{eq: expansion of b}.}
In particular,
\begin{equation}
    f_H^{(N_b)}
    \leq\cdots\leq
    f_H^{(2)}
    \leq
    f_H^{(1)}
    =
    f_H.
\end{equation}

For an angular grid with spacing $\Delta_u$, the steering
term associated with virtual location $v=fp/f_H$ and angular
index $n_\theta$ is
\begin{equation}
    e^{-j2\pi v n_\theta/L_{\Delta_u}},
    \quad
    L_{\Delta_u}
    \triangleq
    \frac{c}{f_H\Delta_u},
    \label{eq: cyclic aperture general spacing}
\end{equation}
where $L_{\Delta_u}$ is the corresponding cyclic-aperture
length. For the finest spacing
$\Delta_u=u_{\mathrm{res}}=c/(Wf_H)$, we have
$L_{\Delta_u}=W$.

However, when $L_{\Delta_u} = W$, for any $i>1$, the largest virtual element generated
by $\mathcal F_i(\epsilon)$ is at most
$Wf_H^{(i)}/f_H$. Hence, the corresponding effective virtual
array has an unavoidable terminal gap
\begin{equation}
    \Delta d_i
    =
    W\left(
        1-\frac{f_H^{(i)}}{f_H}
    \right),
    \quad i>1.
    \label{eq: unavoidable terminal gap}
\end{equation}
This gap cannot be removed by adding physical antennas within
the fixed aperture $[0,W]$.

Therefore, for all effective virtual arrays to be capable of
covering a common cyclic aperture, its length must not exceed
the maximum virtual extent of the lowest-frequency effective
sub-band, i.e.,
\begin{equation}
    L_{\Delta_u}
    =
    \frac{c}{f_H\Delta_u}
    \leq
    W\frac{f_H^{(N_b)}}{f_H}.
    \label{eq: common cyclic aperture condition}
\end{equation}
The finest common angle spacing is obtained by equality,
which yields
\begin{equation}
    \Delta_u
    =
    \frac{c}{Wf_H^{(N_b)}}
    =
    \frac{c}{
        W\left(f_L+B/N_b\right)
    }.
    \label{eq: common angular resolution}
\end{equation}
and 
\[
    L_{\Delta_u} = W\frac{f_H^{(N_b)}}{f_H}.
\]

Under the original virtual coordinate $v=fp/f_H$, the virtual
elements are now interpreted over
$[0,L_{\Delta_u})$. To retain the CC formulation with aperture length $W$, we uniformly rescale the virtual coordinate as
\begin{equation}
    \bar v
    \triangleq
    \frac{W}{L_{\Delta_u}}v
    =
    \frac{f_H}{f_H^{(N_b)}}\frac{fp}{f_H}
    =
    \frac{fp}{f_H^{(N_b)}}.
    \label{eq: normalized virtual coordinate}
\end{equation}
where 
\[
    N_{\theta, \rm FoV} \triangleq \frac{2}{\Delta_u}
\] 
denotes the number of angle pixels for full-FoV angle imaging.
Then, to have full-FoV angle imaging over aperture $[0,W)$, the CC requires
\begin{equation}
    d_{\max,i}
    \leq
    \frac{W}{N_{\theta,\mathrm{FoV}}} = \frac{c}{2f_H^{(N_b)}},
    \quad
    \forall i\in\mathcal I(N_b).
    \label{eq: sufficient dmaxi condition}
\end{equation}

Assume sufficiently dense frequency sampling such that each
active sub-band can be treated as continuous. Then, the normalized virtual elements generated by the $m_a$-th physical antenna over $\mathcal F_i(\epsilon)$ form the cluster
\begin{equation}
    \mathcal C_i(p_{m_a})
    \triangleq
    \left[
        p_{m_a}\frac{f_L^{(i)}}{f_H^{(N_b)}},
        \;
        p_{m_a}\frac{f_H^{(i)}}{f_H^{(N_b)}}
    \right].
    \label{eq: designed virtual array expression}
\end{equation}

A sufficient condition for
\eqref{eq: sufficient dmaxi condition} is
\begin{equation}
    p_{m_a+1}
    \frac{f_L^{(i)}}{f_H^{(N_b)}}
    \leq
    p_{m_a}
    \frac{f_H^{(i)}}{f_H^{(N_b)}}
    +
    \frac{c}{2f_H^{(N_b)}},
    \quad
    \forall i\in\mathcal I(N_b).
    \label{eq: all subband cluster condition}
\end{equation}

To minimize the number of physical antennas, each antenna should be placed as sparsely as permitted by all active frequency sets, namely
\begin{equation}
    p_{m_a+1}
    =
    \min_{i\in\mathcal I(N_b)}
    \left(
        \frac{f_H^{(i)}}{f_L^{(i)}}p_{m_a}
        +
        \frac{c}{2f_L^{(i)}}
    \right),
    \qquad
    p_1=0.
    \label{eq: sparse array recursion min}
\end{equation}
Since $p_{m_a}\ge 0$ and $f_L^{(i)}$ decreases with $i$, the expression inside the minimum is increasing in $i$. Hence, the minimum is attained at $i=1$, and 
\begin{equation}
    p_{m_a+1}
    =
    \frac{f_H}{f_H-B_{\mathrm{eff}}}p_{m_a}
    +
    \frac{c}{2(f_H-B_{\mathrm{eff}})},
    \qquad
    p_1=0.
    \label{eq: sparse array recursion}
\end{equation}
Solving this recursion gives the closed-form antenna locations
in \eqref{eq: non-uniform antenna location_fd}.

The form in \eqref{eq: non-uniform antenna location_fd} is written as the conventional spacing $c/(2f_H)$ multiplied by a dimensionless factor, which makes the sparsity explicit. 
In particular, the term in the square brackets can be much larger than one, indicating that the proposed array can be substantially sparser than the conventional ULA baseline.

Moreover, the design also reveals the effect of channel variation. 
A smaller $N_b$ corresponds to a wider $B_{\rm eff}$ and thus provides more frequency diversity for constructing the virtual array. Hence, fewer physical antennas are required for a prescribed width $W$. 
As $N_b$ increases, more frequency diversity is used to accommodate the channel variation, so a denser physical array is needed to maintain the same imaging capability.

\section{Numerical Results}
\label{sec: simulation}

In this section, we consider four representative frequency bands: C-band ($4$ to $8$ GHz), X-band ($8$ to $12$ GHz), K-band ($21$ to $26$ GHz) and W-band ($77$ to $81$ GHz). For all bands, the frequency spacing is set to $\Delta_f = 40$ MHz. 

In the following condition-number evaluations, we report both the unweighted and weighted system matrices. The weighted condition number is provided in accordance with the deterministic stability analysis, while the unweighted condition number shows the empirical behavior of the original observation model.

\subsection{Coverage Criterion Performance}
We validate the imaging systems constructed by the proposed CC. We first consider the flat-channel case ($N_b = 1$), and then proceed to the frequency-dependent case ($N_b > 1$).
\subsubsection{Flat channel case}
 
We begin with a representative X-band example to verify that the systems generated by the CC are numerically stable under different array geometries.
All arrays occupy the same array width, namely $50\times c/(2f_H)$. Three representative array geometries are considered, and the corresponding outputs are summarized in Table~\ref{table: array patterns}.  
In particular, Array 3 uses only $25$ physical antennas, where $23$ antennas are randomly selected from $\{1,2,\ldots,48\}$ and the two end points $\{0,49\}$ are fixed to preserve the array width. For Array 3, both $N_\theta^\cc$ and condition numbers are averaged over $100$ random realizations.

For all three arrays, the unweighted condition numbers remain below $5$, while the weighted condition numbers remain below $2$, indicating that the CC-designed systems support stable angular-response recovery.

\begin{table}[ht]
\centering
\caption{Array configurations and corresponding system outputs in X-band}
\renewcommand{\arraystretch}{1.2}
\resizebox{\columnwidth}{!}{%
\begin{tabular}{|c|c|c|c|c|c|}
\hline
Array & $M_a$ & \begin{tabular}[c]{@{}c@{}}Geometrical positions \\ (normalized by $c/(2f_H)$)\end{tabular} & \begin{tabular}[c]{@{}c@{}} $N_\theta^\cc$ by\\ \eqref{eq: actual angle pixels minimum} \end{tabular} & $\kappa(\bm{A}^\fl)$ & $\kappa(\widetilde{\bm A}^\fl)$\\ \hline
1     & $50$                                                                     & $\{m_a-1~|~m_a\in\mathcal{I}(50)\}$                                                                               & $50$  & $4.04$ & $1.14$               \\ \hline
2     & $34$                                                                     & \begin{tabular}[c]{@{}c@{}} $\{49(m_a-1)/33~|$\\ $~m_a\in\mathcal{I}(34)\}$\end{tabular}                                                            & $50$  & $4.13$  & $1.53$            \\ \hline
3     & $25$                                                                     & \begin{tabular}[c]{@{}c@{}} 
$\mathcal S\cup\{0,49\}$ \\ 
$\mathcal S\sim\operatorname{Unif}
\{\mathcal T\subseteq\{1,\ldots,48\}: $\\
$|\mathcal T|=23\}$\end{tabular}        & $31$  & $3.91$     & $1.49$        \\ \hline
\end{tabular}%
}
\label{table: array patterns}
\end{table}

\begin{figure}
    \centering
    \includegraphics[width=\linewidth]{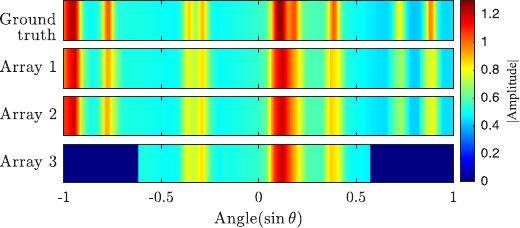}
    \caption{Generated angle images using the array geometries in Table~\ref{table: array patterns}. X-band signal is applied, and SNR = $15$ dB. Image for Array 3 is generated by a representative random array.}
    \label{fig: angle imaging validation-flat channel}
\end{figure}

We then examine the corresponding angle-response recovery.
The ground-truth angular response $\bm \gamma$ is modeled as a smooth, non-sparse, complex-valued function with Gaussian peaks.
Fig.~\ref{fig: angle imaging validation-flat channel} shows the generated angle images, where the horizontal axis corresponds to the angular grid and the color indicates the magnitude of each angle pixel. 
The results show that the systems generated by the CC accurately recover the responses on the prescribed angular grid. 
In particular, Array 2 achieves essentially the same recovery quality as the conventional half-wavelength ULA in Array 1, while using substantially fewer physical antennas. 
Array 3 uses only $25$ antennas with much larger inter-element spacings and still yields stable recovery over its reduced supported FoV.

We next assess the noise robustness of the resulting imaging systems. 
Table~\ref{table: rmse_snr_flat} reports the $\log_{10}$-root-mean-square-error (RMSE) between the recovered and ground-truth $\bm \gamma$ for representative Array 2 configurations under different SNR levels. Each result is averaged over $100$ trials.
For all listed bands, the imaging systems generated by the CC remain numerically stable under noise perturbations. 

\begin{table}[!t]
\centering
\renewcommand{\arraystretch}{1.2}
\caption{RMSE versus SNR. The entries denote the $\log_{10}$-RMSE between the recovered and ground-truth coefficients.}
\label{table: rmse_snr_flat}
\resizebox{!}{!}{%
\begin{tabular}{|c|ccc|}
\hline
\multirow{2}{*}{\textbf{Configuration}} & \multicolumn{3}{c|}{\textbf{SNR}}                                                   \\ \cline{2-4} 
                                         & \multicolumn{1}{l|}{-5 dB} & \multicolumn{1}{l|}{5 dB} & \multicolumn{1}{l|}{15 dB} \\ \hline
C-Array 2 ($\kappa(\bm A^\fl) = 4.52$) & \multicolumn{1}{c|}{-1.21}      & \multicolumn{1}{c|}{-1.72}  & -2.21                       \\ \hline
K-Array 2 ($\kappa(\bm A^\fl) = 3.77$)                              & \multicolumn{1}{c|}{-1.47}    & \multicolumn{1}{c|}{-1.96}   & -2.47                      \\ \hline
W-Array 2 ($\kappa(\bm A^\fl) = 2.31$)                                & \multicolumn{1}{c|}{-1.38}      & \multicolumn{1}{c|}{-1.87}     &  -2.38                          \\ \hline
\end{tabular}%
}
\end{table}

\subsubsection{Frequency-dependent channel case}

We next consider the general case with frequency-dependent channel responses and evaluate the performance of the imaging systems generated by the CC.

We begin with a representative X-band example using Array 2. To model a practically relevant wideband setting, we consider channel responses that vary smoothly with frequency due to frequency-selective reflections of target materials, and set the channel variation order to $N_b = 4$. The threshold in the CC is chosen as $\epsilon = 0.25$. 
Under this setting, the CC yields $N_\theta^\cc=27$. The associated weighted system matrix has condition number $\kappa(\bm A_{\mathrm w}^{\fd})= 89.14$. For reference, the unweighted matrix has condition number
$\kappa(\bm A^{\fd})=69.65$.

\begin{figure}
    \centering
    \includegraphics[width=\linewidth]{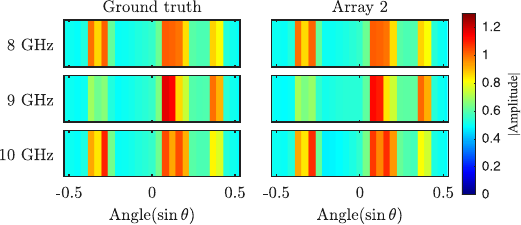}
    \caption{Generated angle images at three frequency samples across the X-band. X-band signal is received by Array 2 defined in Table~\ref{table: array patterns}. The parameters are $N_b = 4$, $\epsilon = 0.25$, and SNR = $15$ dB.}
    \label{fig: freq-dependent imaging}
\end{figure}

Fig.~\ref{fig: freq-dependent imaging} shows the recovered angular responses at three representative frequency samples. Compared with the flat-channel case, the supported number of angle pixels is smaller, since each effective virtual array contains fewer virtual elements. Nevertheless, the recovered responses at all three representative frequency samples remain in close agreement with the ground truth, indicating that the CC still yields a practically reliable imaging system under frequency-dependent channel responses.

\begin{figure}
    \centering
    \includegraphics[width=0.9\linewidth]{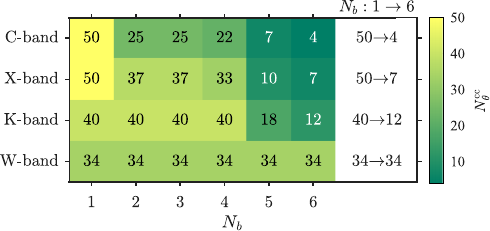}
    \caption{$N_\theta^\cc$ across bands and $N_b$ ($\epsilon=0.25$, Array 2). Each cell shows the exact value, and the rightmost column summarizes the change from $N_b=1$ to $N_b=6$. For all considered configurations, $\log_{10}$-RMSE $<-2$ at an SNR of $15$~dB.}
    \label{fig: Nb vs Ntheta}
\end{figure}

In practice, the value of $N_b$ may vary with the targets and propagation conditions. Therefore, we next examine the effect of the channel-variation complexity $N_b$ on the
imaging performance. 
Fig.~\ref{fig: Nb vs Ntheta} provides $N_\theta^\cc$ for different bands and values of $N_b$.

Several observations can be drawn from Fig.~\ref{fig: Nb vs Ntheta}. 
As $N_b$ increases, $N_\theta^\cc$ generally decreases, showing that stronger
frequency variation reduces the supportable imaging capability. Moreover, the
sensitivity to $N_b$ depends on the value of $\alpha(1) = f_H/B$. For the considered settings, bands
with smaller $\alpha(1)$ support more angle pixels when $N_b$ is small, but their
supported $N_\theta^\cc$ drops more rapidly as $N_b$ increases. 
In contrast, bands with larger $\alpha(1)$ support fewer angle pixels at small $N_b$, but show a weaker degradation as the channel variation complexity grows. 
This suggests that smaller-$\alpha(1)$ bands are preferable when the channel variation is mild, whereas larger-$\alpha(1)$ bands provide a more stable operating regime under stronger frequency selectivity.

\subsection{Non-uniform array design}

We next evaluate the proposed non-uniform array design. As usual, we begin with the flat-channel case with $N_b = 1$, and then consider the general frequency-dependent case with $N_b>1$.

\subsubsection{Flat channel case}
\label{subsubsec: non-uniform array_flat}

To compare the performance of the designed array with different frequency bands, we aim to present the results of full-FoV angle imaging with approximately $100$ angle pixels while using as few physical antennas as possible. To this end, we take the target array width to be $W = 100 \times c/(2f_H)$.

\begin{figure}[t]
    \centering
    \includegraphics[width=\linewidth]{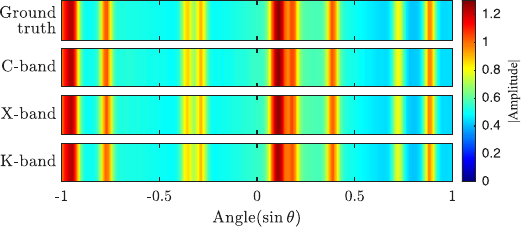}
    \caption{Recovered $\bm{\gamma}$ using the proposed non-uniform arrays in the special case $N_b=1$. In each band, the array is chosen so that the supported full-FoV angle grid is as close as possible to $100$ pixels. SNR = $15$ dB.}
    \label{fig: angle imaging non-uniform array flat}
\end{figure}

Because the closed-form array width in \eqref{eq: array width design} depends on $\alpha(N_b)$ and the number of physical antennas $M_a$ is discrete, the proposed design cannot, in general, realize exactly $100$ angle pixels for every band. We therefore report, for each band, the realizable array design whose supported number of angle pixels is closest to the target value of $100$. The corresponding array settings and resulting system condition numbers are summarized in Table~\ref{tab: nonuniform_flat_array}, and the recovered angular responses are shown in Fig.~\ref{fig: angle imaging non-uniform array flat}.

\begin{table}[t]
\caption{Non-uniform array designs targeting approximately $100$ angle pixels under $N_b=1$}
\centering
\renewcommand{\arraystretch}{1.2}
\resizebox{!}{!}{%
\begin{tabular}{|c|c|c|c|c|c|}
\hline
\textbf{Band} & $\alpha(1) = \frac{f_H}{B}$ & $M_a$ & \begin{tabular}[c]{@{}c@{}}Supported \\ full-FoV $N_\theta$ \end{tabular} & $\kappa(\bm A^\fl)$ & $\kappa(\widetilde{\bm A}^\fl)$ \\ \hline
C-band & $2$ & $7$  & $126$ & $8.72$ & $1.47$ \\ \hline
X-band & $3$ & $10$ & $112$ & $8.90$ & $1.93$\\ \hline
K-band & $5.2$ & $15$ & $99$  & $8.38$ & $2.12$\\ \hline
W-band & $20.25$ & $36$ & $99$  & $5.91$ & $3.05$\\ \hline
\end{tabular}%
}
\label{tab: nonuniform_flat_array}
\end{table}

Table~\ref{tab: nonuniform_flat_array} shows that, although the supported number of angle pixels is not exactly identical across bands, all reported designs yield moderate condition numbers.

Fig.~\ref{fig: angle imaging non-uniform array flat} further shows that all four designs successfully recover the angular response over the full FoV. In addition, as $\alpha(1)$ increases, more physical antennas are required to support a comparable number of full-FoV angle pixels. 
Nevertheless, for all considered bands, the supported number of angle pixels substantially exceeds the number of physical antennas. This clearly contrasts with conventional narrowband array designs, where the recoverable angular dimension is typically on the order of $M_a$.
These results verify that the proposed non-uniform array design can realize full-FoV angle imaging with remarkably few physical antennas while preserving stable recovery.

\subsubsection{Frequency-dependent channel case}

\begin{figure}[t]
\centering
\subfloat[X-band.]{%
    \includegraphics[width=0.49\columnwidth]{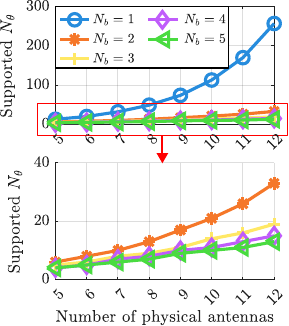}%
    \label{fig: Nb2 Ma vs.Ntheta}%
}
\hfill
\subfloat[K-band.]{%
    \includegraphics[width=0.49\columnwidth]{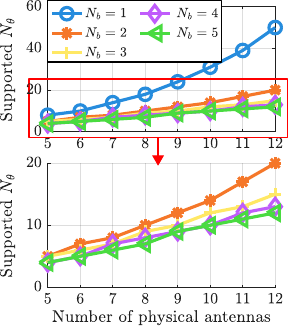}%
    \label{fig: Nb3 Ma vs.Ntheta}%
}
\caption{Supported number of angle pixels versus the number of physical antennas for different $N_b$ with the proposed non-uniform array design.}
\label{fig: Ma vs.Ntheta fd array design}
\end{figure}

We next evaluate the proposed non-uniform array design under frequency-dependent channels. 
For a given antenna budget $M_a$, Fig.~\ref{fig: Ma vs.Ntheta fd array design} shows the number of angle pixels that can be supported in the X- and K-bands under different channel complexities $N_b$. 

Several observations can be made from Fig.~\ref{fig: Ma vs.Ntheta fd array design}. 
First, for both bands, the number of supported angle pixels increases with the number of physical antennas, as a larger array provides more virtual samples for covering the angular grid. 
Second, for a fixed band and a fixed $M_a$, the supported number of angle pixels decreases as $N_b$ increases. 
This is because a larger $N_b$ divides the frequency samples into more active frequency sets, so that each effective virtual array contains fewer elements.
Third, compared with the K-band, the X-band supports more angle pixels under the same $M_a$ and $N_b$, owing to its smaller $\alpha(1)$ and hence wider virtual cluster induced by each physical antenna. 

For all configurations shown in Fig.~\ref{fig: Ma vs.Ntheta fd array design}, the reconstruction RMSE remains below $10^{-2}$, confirming that the resulting operating points remain numerically reliable.

\section{Conclusion}
\label{sec: conclusion}

This paper developed a geometry-driven framework for stable wideband angle imaging with sparse physical arrays. 
By exploiting the frequency-dependent spatial steering of wideband signals, the proposed framework forms a composite virtual aperture from frequency-antenna pairs and supports dense angle imaging over a prescribed FoV. 
We established a wideband angle-imaging model with frequency-dependent channel responses, introduced the CC to characterize the number of stably supportable angle pixels, and proved deterministic condition-number bounds for CC-constructed systems. 
Based on the CC, we further developed a non-uniform sparse-array design that reduces the number of physical antennas while maintaining stable full-FoV recovery. 
Numerical results verified that the proposed CC provides reliable stability guidance and that the designed sparse arrays can support substantially more angle pixels than physical antennas.

\appendix

\subsection{Proof of Theorem~\ref{theorem: cc weighted conditioning flat}}
\label{Proof: cc weighted conditioning flat}

Let $\bm\gamma\in\mathbb C^{N_\theta^\cc}$ be an arbitrary angular response vector supported on the angular grid $\mathcal G_\theta$. Define the spatial-domain response induced by $\bm\gamma$ as
\begin{equation}
    y(v)
    =
    \sum_{n_\theta\in\mathcal G_\theta}
    \gamma(n_\theta)e^{-jv(2\pi n_\theta/W)},
    \qquad v\in[0,W).
    \label{eq: proof aperture response continuous}
\end{equation}
where $v$ is a continuous spatial coordinate. 
The virtual elements $\{v_{m_v}\}_{m_v=1}^{M_v}$ are discrete, generally nonuniform, sampling locations along this coordinate. Hence, the received signal at the $m_v$-th virtual element is the sample
\begin{equation}
    y(v_{m_v})
    =
    \sum_{n_\theta\in\mathcal G_\theta}
    \gamma(n_\theta)e^{-jv_{m_v}(2\pi  n_\theta/W)}
    =
    [\bm A^{\fl}\bm\gamma]_{m_v} .
    \label{eq: proof sampled aperture response}
\end{equation}

Let
\[
    n_{\min}\triangleq \min\mathcal G_\theta,
    \qquad
    n_{\max}\triangleq \max\mathcal G_\theta .
\]
Since $\mathcal G_\theta$ contains $N_\theta^\cc$ consecutive integer indices,
we have
\begin{equation}
    n_{\max}-n_{\min}=N_\theta^\cc-1 .
\end{equation}

Multiplying $y(v)$ by the known unit-modulus modulation
\[
    e^{j2\pi v(n_{\min}+n_{\max})/(2W)}
\]
centers the angular-index range around zero. Hence, after centering, $y(v)$ can be regarded as a bandlimited trigonometric polynomial with respect to $v$, whose relevant one-sided bandwidth is 
\begin{equation}
    \omega_v = \frac{1}{2}\frac{2\pi(n_{\max}-n_{\min})}{W} = 
    \frac{\pi(N_\theta^\cc-1)}{W}.
    \label{eq: proof aperture bandwidth}
\end{equation}

For clarity, the continuous $L^2$ energy over the aperture interval is defined as
\begin{equation}
    \|y\|_{L^2([0,W])}^2
    \triangleq
    \int_0^W |y(v)|^2\,dv .
    \label{eq: proof L2 norm definition}
\end{equation}

The adaptive-weight irregular sampling bound in
\cite[Sec.~4, Th.~6(A)]{feichtinger2021theory} relates this continuous energy to
the Voronoi-weighted sample energy. In the present circular-array setting, the Voronoi weights of virtual elements are given by \eqref{eq: Voronoi weights}, and the sampling-gap
parameter is the maximum wrap-around inter-element gap $d_{\max}$. Therefore, we obtain
\begin{equation}
\begin{aligned}
    \left(1-\frac{d_{\max}\omega_v}{\pi}\right)^2
    \|y\|_{L^2[0,W]}^2
    \le
    \sum_{m_v=1}^{M_v} w_{m_v} |y(v_{m_v})|^2\\
    \le
    \left(1+\frac{d_{\max}\omega_v}{\pi}\right)^2
    \|y\|_{L^2[0,W]}^2.   
\end{aligned}
\label{eq: proof weighted sampling bound angular}
\end{equation}
Using \eqref{eq: proof aperture bandwidth}, define
\begin{equation}
    \rho_\theta
    \triangleq
    \frac{d_{\max}\omega_v}{\pi}
    =
    \frac{(N_\theta^\cc-1)d_{\max}}{W}.
    \label{eq: proof rho theta definition}
\end{equation}

The middle term in \eqref{eq: proof weighted sampling bound angular} can
be written as the energy of the weighted matrix output. Specifically, by
\eqref{eq: proof sampled aperture response},
\begin{equation}
    \sum_{m_v=1}^{M_v} w_{m_v} |y(v_{m_v})|^2
    =
    \left\|
    \bm W_v^{1/2}\bm A^{\fl}\bm\gamma
    \right\|_2^2
    =
    \left\|
    \widetilde{\bm A}^{\fl}\bm\gamma
    \right\|_2^2 .
    \label{eq: proof weighted matrix energy}
\end{equation}

Moreover, by the orthogonality of complex exponentials over $[0,W)$, we have
\begin{equation}
    \int_0^W
    e^{-j2\pi v(n_\theta-n_\theta')/W}\,dv
    =
    W\delta_{n_\theta,n_\theta'},
\end{equation}
and therefore
\begin{equation}
    \|y\|_{L^2[0,W]}^2
    =
    W\|\bm\gamma\|_2^2 .
    \label{eq: proof parseval}
\end{equation}

Combining \eqref{eq: proof weighted sampling bound angular},
\eqref{eq: proof weighted matrix energy}, and \eqref{eq: proof parseval} yields
\begin{equation}
    W(1-\rho_\theta)^2\|\bm\gamma\|_2^2
    \le
    \left\|
    \widetilde{\bm A}^{\fl}\bm\gamma
    \right\|_2^2
    \le
    W(1+\rho_\theta)^2\|\bm\gamma\|_2^2 .
    \label{eq: proof weighted frame inequality}
\end{equation}
Taking the infimum and supremum of
\eqref{eq: proof weighted frame inequality} over all unit-norm
$\bm\gamma$ gives
\begin{equation}
    \sigma_{\min}\!\left(\widetilde{\bm A}^{\fl}\right)
    \ge
    \sqrt W(1-\rho_\theta),
    \quad
    \sigma_{\max}\!\left(\widetilde{\bm A}^{\fl}\right)
    \le
    \sqrt W(1+\rho_\theta).
\end{equation}
Therefore,
\begin{equation}
    \kappa\!\left(\widetilde{\bm A}^{\fl}\right)
    \le
    \frac{1+\rho_\theta}{1-\rho_\theta}.
    \label{eq: proof of A_tilde bound}
\end{equation}

Finally, by the CC,
\begin{equation}
    N_\theta^\cc
    \le
    \left\lfloor \frac{W}{d_{\max}}\right\rfloor
    \le
    \frac{W}{d_{\max}} .
\end{equation}
Hence,
\begin{equation}
    \rho_\theta
    =
    \frac{(N_\theta^\cc-1)d_{\max}}{W}
    \le
    1-\frac{d_{\max}}{W}
    <1 .
    \label{eq: rho_theta bound}
\end{equation}
Since $(1+\rho)/(1-\rho)$ is monotonically increasing for $0\le\rho<1$, applying
\eqref{eq: rho_theta bound} to \eqref{eq: proof of A_tilde bound} gives
\begin{equation}
    \kappa\!\left(\widetilde{\bm A}^{\fl}\right)
    \le
    \frac{1+(1-d_{\max}/W)}{1-(1-d_{\max}/W)}
    =
    2\frac{W}{d_{\max}}-1 .
\end{equation}
This completes the proof.

\subsection{Proof of Lemma~\ref{lem: cond of BiA}}
\label{proof: cond of DiA}

Let $\Omega_i$ denote the row indices associated with the active frequency
set $\mathcal F_i(\epsilon)$, and let $\Omega_i^c$ denote its complement.
Let $\widetilde{\bm A}_i$ and $\widetilde{\bm A}_i^\perp$ be the row submatrices of
$\widetilde{\bm A}^{\fl}$ formed by selecting the rows in $\Omega_i$ and
$\Omega_i^c$, respectively. Then
\begin{equation}
    (\widetilde{\bm A}^{\fl})^\H\widetilde{\bm A}^{\fl}
    =
    \widetilde{\bm A}_i^\H\widetilde{\bm A}_i
    +
    (\widetilde{\bm A}_i^\perp)^\H\widetilde{\bm A}_i^\perp .
    \label{eq:Afl_weighted_decomp_appendix}
\end{equation}

We first justify that $\widetilde{\bm A}_i$ is full column rank. By the CC, the effective
virtual array $\mathcal V_i^\epsilon$ supports $N_\theta^\cc$ consecutive
angle pixels, so the corresponding active steering matrix satisfies the
same sampling-gap condition used in Theorem~\ref{theorem: cc weighted conditioning flat}.
Applying that theorem to the active virtual array gives a strictly positive
lower singular-value bound for a weighted active
matrix. Since row scalings by positive weights do not change column rank, the active row submatrix $\widetilde{\bm A}_i$ is full column rank.

Since $\bm D_i$ is diagonal over the frequency-dependent rows, the Gram
matrix of $\bm D_i\widetilde{\bm A}^{\fl}$ can be written as
\begin{equation}
\begin{aligned}
    (\bm D_i\widetilde{\bm A}^{\fl})^\H
    (\bm D_i\widetilde{\bm A}^{\fl})
    &=
    \widetilde{\bm A}_i^\H \bm \Lambda_i \widetilde{\bm A}_i
    +
    (\widetilde{\bm A}_i^\perp)^\H \bm \Lambda_i^\perp \widetilde{\bm A}_i^\perp,
\end{aligned}
\label{eq:DiA_weighted_gram_decomp_appendix}
\end{equation}
where $\bm \Lambda_i$ and $\bm \Lambda_i^\perp$ are diagonal matrices
collecting $|\beta_{i,m_f}|^2$ on the active and inactive rows, respectively.
By definition of $\mathcal F_i(\epsilon)$,
$|\beta_{i,m_f}|>\epsilon$ on active rows and
$|\beta_{i,m_f}|\le \epsilon$ on inactive rows. 
Hence,
\begin{equation}
    \epsilon^2 \bm I \preceq \bm \Lambda_i \preceq \bm I,
    \qquad
    \bm 0 \preceq \bm \Lambda_i^\perp \preceq \epsilon^2\bm I .
    \label{eq:lambda_weighted_bounds_appendix}
\end{equation}

From \eqref{eq:DiA_weighted_gram_decomp_appendix} and
\eqref{eq:lambda_weighted_bounds_appendix}, we have
\begin{equation}
    (\bm D_i\widetilde{\bm A}^{\fl})^\H
    (\bm D_i\widetilde{\bm A}^{\fl})
    \succeq
    \epsilon^2\widetilde{\bm A}_i^\H\widetilde{\bm A}_i ,
\end{equation}
which gives
\begin{equation}
    \sigma_{\min}^2(\bm D_i\widetilde{\bm A}^{\fl})
    \ge
    \epsilon^2\sigma_{\min}^2(\widetilde{\bm A}_i).
    \label{eq:sigma_min_weighted_DiA_appendix}
\end{equation}
Similarly,
\begin{equation}
    (\bm D_i\widetilde{\bm A}^{\fl})^\H
    (\bm D_i\widetilde{\bm A}^{\fl})
    \preceq
    \widetilde{\bm A}_i^\H\widetilde{\bm A}_i
    +
    \epsilon^2(\widetilde{\bm A}_i^\perp)^\H\widetilde{\bm A}_i^\perp,
\end{equation}
and therefore
\begin{equation}
    \sigma_{\max}^2(\bm D_i\widetilde{\bm A}^{\fl})
    \le
    \sigma_{\max}^2(\widetilde{\bm A}_i)
    +
    \epsilon^2\sigma_{\max}^2(\widetilde{\bm A}_i^\perp).
    \label{eq:sigma_max_weighted_DiA_intermediate_appendix}
\end{equation}

It remains to bound $\sigma_{\max}(\widetilde{\bm A}_i^\perp)$. From
\eqref{eq:Afl_weighted_decomp_appendix},
\begin{equation}
    (\widetilde{\bm A}_i^\perp)^\H\widetilde{\bm A}_i^\perp
    =
    (\widetilde{\bm A}^{\fl})^\H\widetilde{\bm A}^{\fl}
    -
    \widetilde{\bm A}_i^\H\widetilde{\bm A}_i .
\end{equation}
Since $\widetilde{\bm A}_i$ is full column rank,
$\widetilde{\bm A}_i^\H\widetilde{\bm A}_i\succeq\sigma_{\min}^2(\widetilde{\bm A}_i)\bm I$. It follows that
\begin{equation}
    (\widetilde{\bm A}_i^\perp)^\H\widetilde{\bm A}_i^\perp
    \preceq
    (\widetilde{\bm A}^{\fl})^\H\widetilde{\bm A}^{\fl}
    -
    \sigma_{\min}^2(\widetilde{\bm A}_i)\bm I .
\end{equation}
Thus,
\begin{equation}
    \sigma_{\max}^2(\widetilde{\bm A}_i^\perp)
    \le
    \sigma_{\max}^2(\widetilde{\bm A}^{\fl})
    -
    \sigma_{\min}^2(\widetilde{\bm A}_i).
    \label{eq:sigma_weighted_Aiperp_bound_appendix}
\end{equation}
Substituting \eqref{eq:sigma_weighted_Aiperp_bound_appendix} into
\eqref{eq:sigma_max_weighted_DiA_intermediate_appendix} yields
\begin{equation}
\begin{aligned}
    \sigma_{\max}^2(\bm D_i\widetilde{\bm A}^{\fl})
    \le\;
    &\sigma_{\max}^2(\widetilde{\bm A}_i)
    +
    \epsilon^2
    \left(
    \sigma_{\max}^2(\widetilde{\bm A}^{\fl})
    -
    \sigma_{\min}^2(\widetilde{\bm A}_i)
    \right).
\end{aligned}
\end{equation}
Together with \eqref{eq:sigma_min_weighted_DiA_appendix}, this proves the lemma.

\subsection{Proof of Theorem~\ref{theorem: bound of A_fd}}
\label{proof: bound of A_fd}

Let
\begin{equation}
\begin{aligned}
    \eta_{\min}
    &\triangleq
    \min_{i\in\mathcal I(N_b)}
    \sigma_{\min}^2(\bm D_i\widetilde{\bm A}^{\fl}),\\
    \eta_{\max}
    &\triangleq
    \max_{i\in\mathcal I(N_b)}
    \sigma_{\max}^2(\bm D_i\widetilde{\bm A}^{\fl}).
\end{aligned}
\end{equation}
Since $\bm Q_D$ is block diagonal, its extreme eigenvalues are
\begin{equation}
    \lambda_{\min}(\bm Q_D)=\eta_{\min},
    \qquad
    \lambda_{\max}(\bm Q_D)=\eta_{\max}.
\end{equation}
By definition,
\begin{equation}
    (\widetilde{\bm A}_{\mathrm w}^{\fd})^\H
    \widetilde{\bm A}_{\mathrm w}^{\fd}
    =
    \bm Q_D+\bm Q_R,
    \qquad
    \|\bm Q_R\|_2=\delta .
\end{equation}
Applying Weyl's inequality gives
\begin{equation}
    \lambda_{\min}\!\left(
    (\widetilde{\bm A}_{\mathrm w}^{\fd})^\H
    \widetilde{\bm A}_{\mathrm w}^{\fd}
    \right)
    \ge
    \eta_{\min}-\delta,
\end{equation}
and
\begin{equation}
    \lambda_{\max}\!\left(
    (\widetilde{\bm A}_{\mathrm w}^{\fd})^\H
    \widetilde{\bm A}_{\mathrm w}^{\fd}
    \right)
    \le
    \eta_{\max}+\delta .
\end{equation}
Under the condition $\delta<\eta_{\min}$, the Gram matrix is positive
definite. Therefore,
\begin{equation}
    \kappa^2(\widetilde{\bm A}_{\mathrm w}^{\fd})
    \le
    \frac{\eta_{\max}+\delta}{\eta_{\min}-\delta}.
\end{equation}
Recall from \eqref{eq: weighted fd block form} that
\[
    \bm A_{\mathrm w}^{\fd}
    =
    \widetilde{\bm A}_{\mathrm w}^{\fd}
    (\bm E\otimes\bm I_{N_\theta}).
\]
The DFT-row matrix $\bm E$ is scaled unitary, i.e.,
$\bm E\bm E^\H=N_b\bm I_{N_b}$. Hence
$\bm E\otimes\bm I_{N_\theta}$ scales all singular values by the same factor
$\sqrt{N_b}$ and does not change the condition number. Thus,
$\kappa(\bm A_{\mathrm w}^{\fd})
=\kappa(\widetilde{\bm A}_{\mathrm w}^{\fd})$, and
\begin{equation}
    \kappa(\bm A_{\mathrm w}^{\fd})
    \le
    \sqrt{
    \frac{
    \max_{i\in\mathcal I(N_b)}
    \sigma_{\max}^2(\bm D_i\widetilde{\bm A}^{\fl})+\delta
    }{
    \min_{i\in\mathcal I(N_b)}
    \sigma_{\min}^2(\bm D_i\widetilde{\bm A}^{\fl})-\delta
    }
    }.
\end{equation}
This completes the proof.

\bibliographystyle{IEEEtran}
\bibliography{refs}

\end{document}